\documentclass[fp,letterpaper,,twocolumn]{jpsj3}
\usepackage{txfonts}
\usepackage{amsmath,amssymb}
\usepackage{comment}
\newcommand{\means}[1]{\langle#1\rangle}
\ifx\pdfoutput\undefined
\usepackage[dvipdfmx]{graphicx}
\usepackage[dvipdfmx]{color}
\else
\usepackage{graphicx}
\usepackage{color}
\fi

\title{
  Symmetry breaking states in the half-filled two-orbital Hubbard model with crystalline electric field
}
\author{Kosuke Ishigaki, Joji Nasu, and Akihisa Koga}
\inst{
  Department of Physics, Tokyo Institute of Technology,
  Meguro, Tokyo 152- 8551, Japan
}

\abst{
  We investigate the half-filled two-orbital Hubbard model with the crystalline electric field
  using dynamical mean-field theory combined with the continuous-time quantum
  Monte Carlo simulations.
  We systematically study how the interplay of the intra- and interorbital Coulomb
  interations together with the Hund coupling realizes the diagonal and off-diagonal ordered states.
  It is found that the antiferroorbital ordered state is realized in the Hubbard model,
  in addition to the antiferromagnetically ordered and excitonic states.
  The competition between the antiferroorbital ordered and excitonic states
  close to the band insulating state is addressed.
}
\begin{document}
\maketitle

\section{Introduction}
Strongly correlated electron systems with orbital degrees of freedom
have attracted much interest.
One of the exotic phenomena is the excitonic insulator~\cite{Mott,Jerome,HalperinRice}.
This state is defined by the emergence of the condensation of excitons, which are composed of pairs of electron in the conduction band and
hole in the valence band.
Recently, the chalcogenites $\mathrm{Ta_2NiSe_5}$\cite{PhysRevLett.103.026402,wakisaka2012photoemission,kaneko2013orthorhombic} and $1T$-TiSe$_2$~\cite{ishioka2010chiral,watanabe2015charge,porer2014non} have been synthesized.
The former material exhibits the flattening of the band structure around the Fermi level
with decreasing temperature, which is considered to be manifestation of
the excitonic insulator~\cite{Seki_2014,wakisaka2012photoemission}, and
an attempt to control the magnitude of the excitation gap has been performed using the pump-probe spectroscopy~\cite{Mor}.
Moreover, the pressure-induced superconductivity has been observed at low temperatures.
In the compound $1T$-TiSe$_2$,
the charge-density-wave (CDW) transition has been observed~\cite{Salvo1976,Zunger1978}.
It has been suggested that the presence of an excitonic insulator phase has
a major effect on the transition~\cite{Cercellier2007},
which attracts much interest
on the chalcogenites~\cite{yamada2016fflo,Sugimoto_2018,Matsuura_2016,Seki_2014}.
In addition to them, the cobaltate $\rm{Pr_{0.5}Ca_{0.5}CoO_3}$ is also proposed
as a candidate of the excitonic insulator owing to the presence of the spin-state degree of
freedom inherent to the cobalt ions~\cite{kunevs2014excitonic,kunevs2014condens,tsubouchi2002simultaneous,fujita2004transport,nasu2016phase,Tatsuno2016}.
These stimulate further theoretical and experimental investigations
on the excitonic state.

The simple model to describe the excitonic insulator should be
the two-orbital Hubbard model with the crystalline electric field.
In the model, the excitonic state is characterized
by the spontaneous hybridization between two orbitals.
The ground state properties in the two-orbital model have been addressed
by means of the variational cluster
approximations~\cite{PhysRevB.85.165135,kaneko2014roles,kaneko2015competition,Hamada}
and dynamical mean-field theory~\cite{kunevs2014excitonic,kunevs2014condens,kunevs2014phase},
where the excitonic state is widely realized between the antiferromagnetically (AFM) ordered
and band insulating (BI) states.
The role of the Hund coupling, pair hoppings, and crystalline electric field
for the excitonic insulator has been examined, where
various low temperature states appear such as
excitonic state together with the ferromagnetic, antiferromagnetic, or
nematic order~\cite{kaneko2014roles,kaneko2015competition,kunevs2014excitonic,kunevs2014condens,kunevs2014phase}.
Also, a Fulde-Ferrell-Larkin-Ovchinnikov-type excitonic insulating state,
which is characterized by the condensation of
excitons with finite center-of-mass momentum,
has been suggested~\cite{yamada2016fflo,Sugimoto_2018}.
Most of theoretical work have considered the multiorbital system
with large intraorbital Coulomb interactions.

On the other hand, in certain compounds, the interorbital interaction
is effectively larger than the intraorbital interaction.
For example, in the fullerene-based solids
$A_3\rm C_{60}$ ($A$=alkali metal)~\cite{Fullerene1,Fullerene2,Fullerene3},
the electron-phonon coupling
should yield the large interorbital interaction and antiferromagnetic Hund coupling
in the triply-degenerate $t_{1u}$ orbitals,
where interesting low temperature properties have been
discussed~\cite{Ishigaki1,Ishigaki2,PhysRevLett.118.177002,nomura2015unified,capone2002strongly,PhysRevB.91.085108,Misawa2017pre}.
However, it remains unclear whether or not the excitonic state is stable against the large interorbital interactions.

In this paper, to reveal the effects of the spin and orbital fluctuations
due to the local Coulomb interactions, we investigate the two-orbital Hubbard model
with a crystalline field using dynamical mean-field theory (DMFT)~\cite{DMFT1,DMFT2,DMFT3}
combined with the continuous-time quantum Monte Carlo (CTQMC) simulations~\cite{CTQMC,CTQMCREV}.
When the intra- and inter-orbital Coulomb interactions are nearly equal,
AFM and antiferroorbital (AFO) orders are suppressed, and
both spin and orbital fluctuations are enhanced.
We find an excitonic phase at low temperatures,
which appears attributed to the enhancement of these fluctuations.
We also clarify that the Hund coupling stabilizes the excitonic state near the AFM,
while little affects the AFO state.

The paper is organized as follows.
In Sec.~\ref{sec:model},
we introduce the two-orbital Hubbard model and briefly summarize
the framework of DMFT~\cite{DMFT1,DMFT2,DMFT3}.
In Sec.~\ref{sec:results},
we study how stable the excitonic state is at low temperatures.
The effect of the Hund coupling is addressed in Sec.~\ref{sec:Hund}.
A summary is given in the final section.

\section{Model and Method}\label{sec:model}
We here introduce a two-orbital Hubbard model with a crystalline field as a simple model
to describe the excitonic state.
This is given by the following Hamiltonian as,
\begin{align}\label{2orb}
   \mathcal{H}= &\sum_{\means{i,j}\alpha\sigma}tc^\dagger_{i\alpha\sigma}c_{j\alpha\sigma}
   +\frac{\Delta_{\mathrm{CF}}}{2}\sum_{i\sigma}(n^a_{i\sigma}-n^b_{i\sigma}) \nonumber \\
   & +U\sum_{i\alpha}n^\alpha_{i\uparrow}n^\alpha_{i\downarrow}
      +U'\sum_{i\sigma}n^a_{i\sigma}n^b_{i-\sigma}
      +(U'-J)\sum_{i\sigma}n^a_{i\sigma}n^b_{i\sigma},
\end{align}
where $c_{i\alpha\sigma}$ is an annihilation operator for an electron
with spin $\sigma(=\uparrow, \downarrow)$ and orbital index $\alpha(=a,b)$ at the $i$th site
and $n^{\alpha}_{i\sigma}=c^\dagger_{i\alpha\sigma}c_{i\alpha\sigma}$.
$t$ is the transfer integral between nearest neighbor sites
for the $\alpha$th orbital, and
$\Delta_{\mathrm{CF}}$ is the band splitting due to the crystalline electric field.
$U (U')$ is the intraorbital (interorbital) interaction and
$J$ is the Hund coupling.

In this paper, we consider the infinite-dimensional Hubbard model to examine
the stability of the excitonic phase~\cite{kunevs2014excitonic,kunevs2014phase,Kim_2017}.
To this end, we employ DMFT~\cite{DMFT1,DMFT2,DMFT3},
where local electron correlations leading to the excitonic state are taken into account properly.
In DMFT, the lattice model is mapped to the problem
of a single impurity model connected dynamically to an effective medium.
The lattice Green's function is obtained via the self-consistency conditions
imposed on the impurity problem.
The treatment is exact in the limit of the infinite dimensions, and even in three dimensions,
DMFT has explained a lot of phenomena emergent in strongly correlated systems.
In fact, DMFT has been applied to multiorbital systems and successfully described interesting low temperature phenomena such as
correlated metallic state~\cite{Han,Imai,Koga2band1,Koga2band2},
Mott transitions~\cite{Rozenberg,Ono_2001,Ohashi_2001,OSMT1,OSMT2,PhysRevLett.118.177002},
magnetism~\cite{Momoi,Held,Ishigaki2},
superconductivity~\cite{PhysRevB.91.085108,Ishigaki1,Hoshino_Werner_2016} and
excitonic state~\cite{kunevs2014excitonic,kunevs2014phase,Kim_2017}.

The excitonic state is characterized by the spontaneous mixing of two orbitals,
and we need to consider the off-diagonal Green's functions in the orbital basis.
The Green's function should be represented
by the 2-by-2 matrix~\cite{Georges_1993} as,
\begin{align}
 \hat{G}(\tau)_{\sigma\sigma'} = \left(
 \begin{array}{cc}
    G_{a\sigma}(\tau) & F_{\sigma\sigma'}(\tau) \\
    F^*_{\sigma'\sigma}(\tau) & G_{b\sigma'}(\tau)
 \end{array}
 \right),
\end{align}
with
\begin{eqnarray}
 G_{\alpha\sigma}(\tau)  &=& \means{T_\tau[c^\dagger_{\alpha\sigma}(\tau)c_{\alpha\sigma}(0)]}, \\
 F_{\sigma\sigma'}(\tau) &=&  \means{T_\tau[c^{\dagger}_{a\sigma}(\tau)c_{b\sigma'}(0)]},\\
 F^*_{\sigma\sigma'}(\tau) &=&  \means{T_\tau[c^{\dagger}_{b\sigma}(\tau)c_{a\sigma'}(0)]},
\end{eqnarray}
where $T_\tau$ is the imaginary-time ordered operator.
Note that $F$ and $F^*$ become nonzero in the excitonic phase.
In DMFT, the lattice Green's function is given by the site-diagonal selfenergy $\hat{\Sigma}$ as
\begin{eqnarray}
  \hat{G}^{-1}(k,i\omega_n)=(i\omega_n+\mu-\epsilon_k)\hat{\sigma}_0 -\hat{\Sigma}(i\omega_n),
\end{eqnarray}
where $\mu$ is the chemical potential,
$\omega_n[=(2n+1)\pi T]$ is the Matsubara frequency, and $T$ is the temperature,
and $\hat{\sigma}_0$ is the identity matrix.
$\epsilon_k$ is the dispersion relation of each orbital for noninteracting system
and
$\hat{\Sigma}(i\omega_n)$ is the local selfenergy.
The local Green's function is obtained as,
\begin{eqnarray}
\hat{G}_{\rm loc}(i\omega_n)=\int dk \hat{G}(k,i\omega_n).
\end{eqnarray}
In the paper, we use a semicircular density of state,
\begin{eqnarray}
  \rho(x)=\frac{2}{\pi D}\sqrt{1-\left(\frac{x}{D}\right)^2},
\end{eqnarray}
which corresponds to an infinite coordination Bethe lattice,
where $D$ is the half bandwidth.
The self-consistency equation is then given by
\begin{eqnarray}
  \hat{\cal G}_\sigma^{-1}(i\omega_n)=(i\omega_n+\mu)\hat{\sigma}_0-\frac{D^2}{4}\hat{G}_{{\rm loc},\sigma}(i\omega_n),
\end{eqnarray}
where $\hat{\cal G}_\sigma(i\omega_n)$ is the effective medium for the effective impurity model.
We iterate the self-consistency equations
until the desired numerical accuracy is achieved.
In our study, we use, as an impurity solver,
the hybridization-expansion continuous-time quantum Monte Carlo simulations~\cite{CTQMC,CTQMCREV},
which is one of the most powerful methods to discuss finite-temperature properties
in the multiorbital model.

Now, we define the order parameters for the possible states
in the Hubbard model on the bipartite lattice.
The order parameters for the antiferromagnetic (AFM) and antiferroorbital (AFO)
ordered states are given as
\begin{eqnarray}
  m_{AFM}&=&\frac{1}{N}\sum_{i\alpha} \left|(-1)^i\left(\means{n_{i\uparrow}^\alpha}-\means{n_{i\downarrow}^\alpha}\right)\right|,\\
  m_{AFO}&=&\frac{1}{N}\sum_{i\sigma}(-1)^i\left(\means{n_{i\sigma}^a}-\means{n_{i\sigma}^b}\right).
\end{eqnarray}
The order parameter for the excitonic state is given by the off-diagonal Green's functions as
\begin{align}
   \phi_{\sigma\sigma'} = \means{c^\dagger_{a\sigma} c_{b\sigma'} }
  =\lim_{\tau\to0+} F_{\sigma\sigma'}(\tau),
\end{align}
where $\phi_{\sigma\sigma'}$ characterizes the excitonic phase associated with its spin state.
In general, $\phi_{\sigma\sigma'}$ is a complex quantity
originating from the relative phase between $a$ and $b$ orbital states.
The Hamiltonian Eq.~(\ref{2orb}) is invariant under the transformation $c_{ia\sigma}\to c_{ia\sigma} e^{i\varphi}$, namely, it possesses the U(1) symmetry around the $\tau^z (=n^a-n^b)$ axis.
Thus, we can choose the phase of $\phi_{\sigma\sigma'}$ so that it is real.
Note that this procedure is not correct in the presence of the pair hopping, which violates the U(1) orbital symmetry~\cite{kunevs2014phase}.
  In addition, as we neglect the spin exchange in the Hund coupling, the SU(2) symmetry is absent in the spin space.
Instead, the U(1) spin symmetry along $S^z$ axis is present in the Hamiltonian Eq.~(\ref{2orb}).
Due to the U(1) symmetry along $S^z$ axis,
the excitonic orders with parallel and antiparallel spins are doubly degenerate.
Therefore, in the present system,
we can assume $\phi\equiv \phi_{\uparrow\uparrow}=\phi_{\downarrow\downarrow}$ for the parallel-spin excitonic state and $\bar{\phi}\equiv \phi_{\uparrow\downarrow}=\phi_{\downarrow\uparrow}$ for the antiparallel-spin excitonic state by fixing the their phase factors.

\section{Results}\label{sec:results}
\begin{figure}[tb]
  \begin{center}
    \includegraphics[width=8cm]{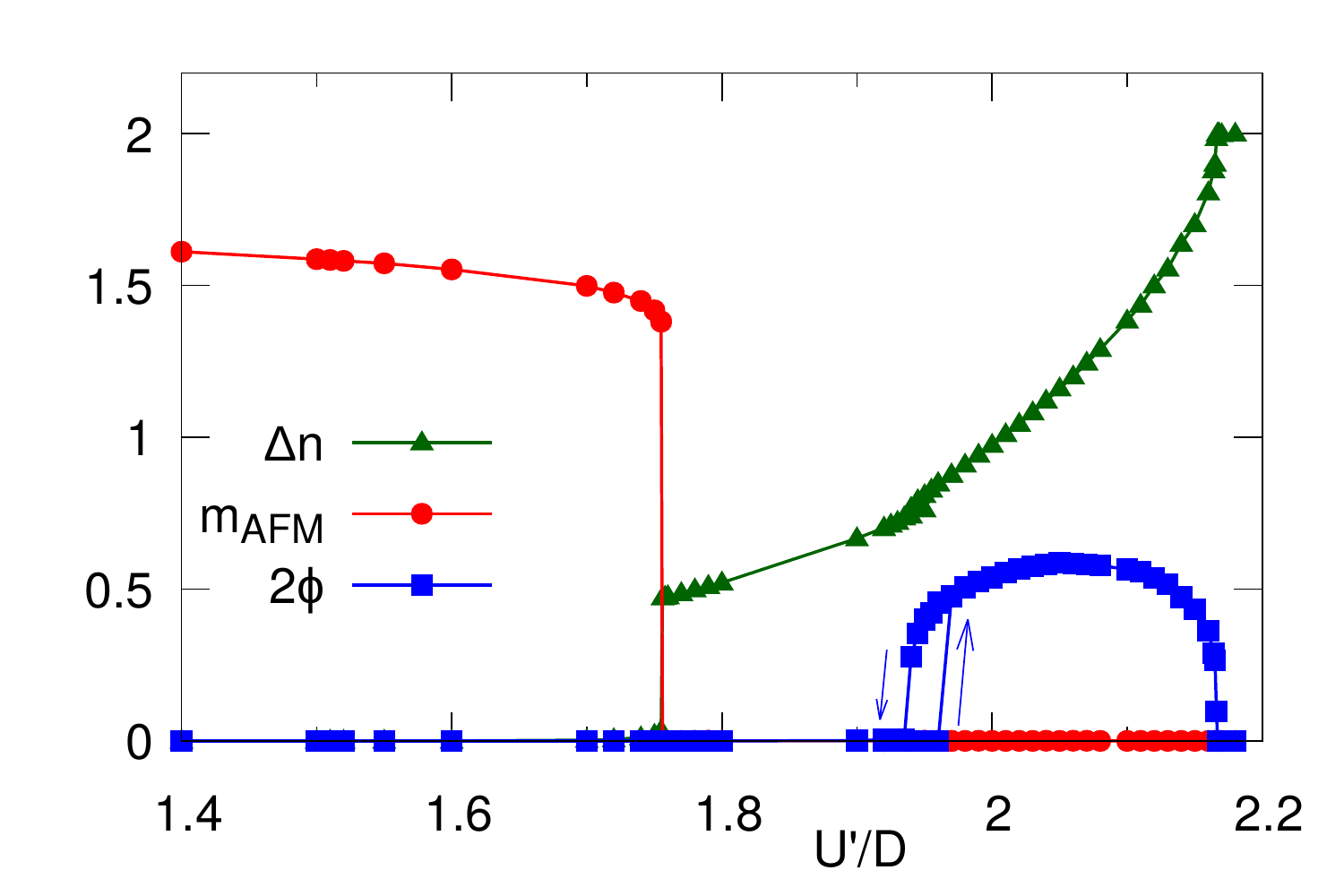}
    \caption{Orbital moment $\Delta n$, order parameter of the excitonic state $\phi$ and
      antiferromagnetically ordered state $m_M$ in the two-orbital system
      with $U/D=2.0$ and $\Delta_{\mathrm{CF}}/D=0.3$ at the temperature $T/D=0.02$.
 }
 \label{ExD0.3}
 \end{center}
 \end{figure}

We start with the two-orbital system without the Hund coupling.
Treating the intraorbital interaction $U$ and interorbital one $U'$ as free parameters,
we clarify the role of these interactions in realizing the excitonic state at low
temperatures~\cite{PhysRevB.85.165135,kaneko2014roles,kaneko2015competition,kunevs2014excitonic,kunevs2014condens,kunevs2014phase}.
Note that, in this simple case,
the excitonic states do not depend on the spins, namely,
$\phi=\bar{\phi}$.
Calculating electron number and order parameters $m_{AFO}, m_{AFM}$, and $\phi$,
we obtain the results in the system with
$U/D=2.0$ and $\Delta_{\mathrm{CF}}/D=0.3$,
as shown in Fig.~\ref{ExD0.3}.
When $U'/D<1.76$,
the intraorbital interaction $U$ is dominant while
the band splitting $\Delta_{\rm CF}$ is irrelevant in the system.
Therefore, the electron numbers in the two orbitals are almost the same,
and the local orbital disproportion is suppressed as $\Delta n\sim 0$, where $\Delta n$ is defined by $\Delta n=\frac{1}{N}\sum_{i\sigma}\left( \means{n_{i\sigma}^a} - \means{n_{i\sigma}^b} \right)$.
The AFM is then realized with the staggered magnetization $m_{AFM}$.
Note that the interorbital Coulomb interaction 
hardly affects
the staggered magnetization.

Increasing the interorbital interaction $U'$,
the staggered magnetization suddenly vanishes and the difference of orbital occupations is largely enhanced as
$\Delta n\sim 0.5$ at $U'/D\sim 1.76$.
This implies that the first-order phase transition occurs to the metallic state although we do not find the hysteresis.
Further increase of the interorbital interaction $U'$ monotonically increases
the orbital moment.
Around $U'/D\sim 1.97$, a tiny jump singularity appears
in the electron number and $\Delta n$,
where the order parameter $\phi$ is suddenly induced.
This indicates the appearance of an excitonic state via the first-order phase transition accompanied by the hysteresis.
With increasing the interorbital interaction $U'$, the order parameter $\phi$ increases.
However, further increase of $U'$ suppresses the order parameter $\phi$.
Finally, this vanishes and the band insulating phase with the lower band $b$ fully occupied appears at $U'/D=2.16$.
In the present case with $\Delta_{\rm CF}/D=0.3$, we have confirmed that
the excitonic state is realized close to the band insulating state and
the interorbital interaction $U'$
does not lead to drastic change around the symmetric condition $U'/D=U/D=2$.

\begin{figure}[tb]
  \centering
  \includegraphics[width=8cm]{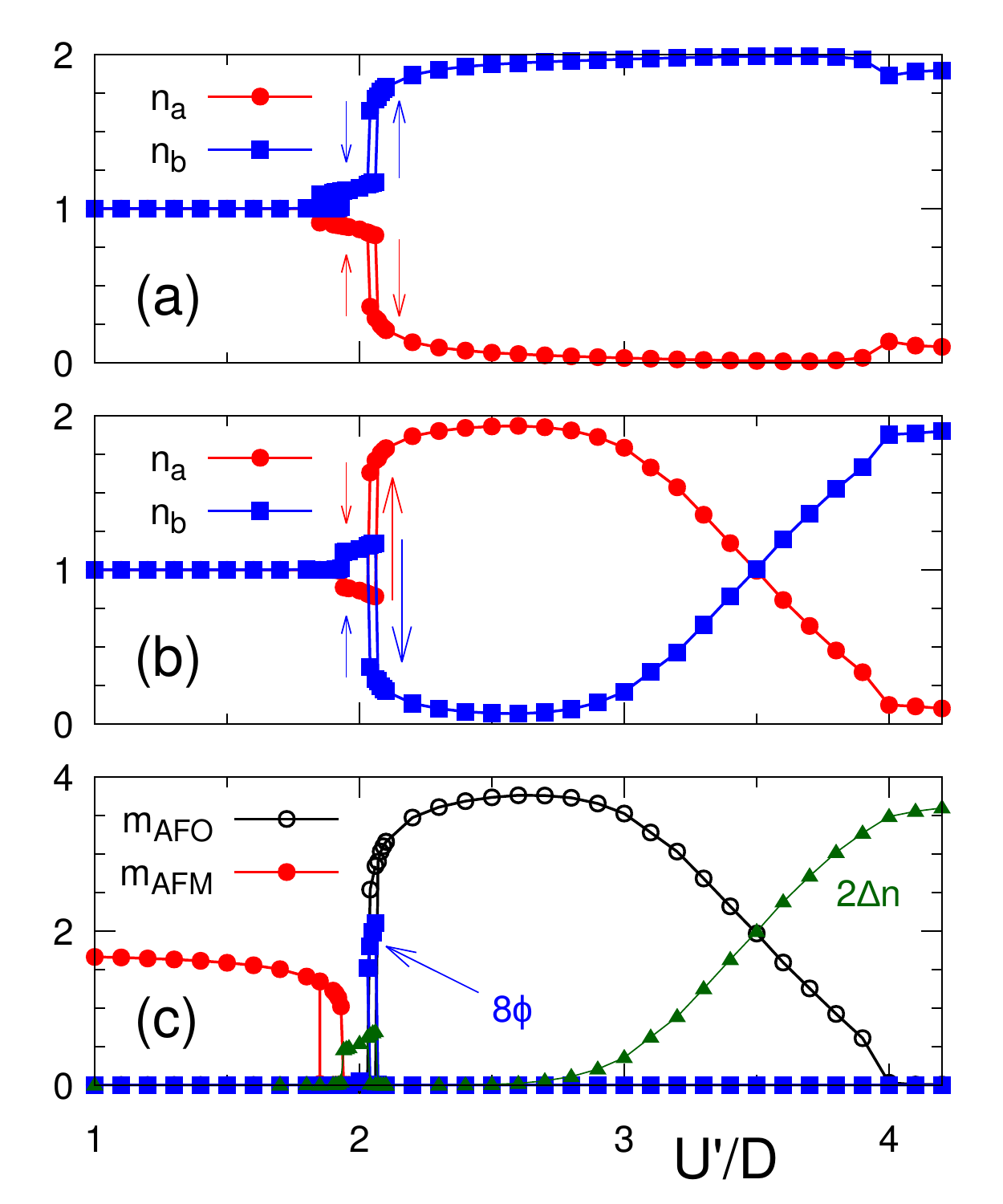}
  \caption{
    Physical quantities as a function of $U'/D$ in the system with
    $\Delta_{\mathrm{CF}}=0.1$ and $U/D=2.0$ at the temperature $T/D=0.02$.
    (a) and (b) represent number of electrons in the sublattices 1 and 2, respectively.
    (c) shows the staggered magnetization and order parameters for excitonic and antiferro-orbital states.
  }
  \label{AFO}
\end{figure}

Figure~\ref{AFO} shows the electron number in both sublattices (1 and 2) and physical quantities
at smaller $\Delta_{\mathrm{CF}}/D=0.1$.
When $U'<U(=2D)$, the AFM, metallic, and excitonic states are realized,
which is similar behavior to the larger $\Delta_{\rm CF}$ case, as discussed above.
By contrast, when the interorbital interaction is larger than the intraorbital interaction
$(2<U'/D<4)$, different behavior appears, where the staggered orbital moment $m_{AFO}$ is finite
and the AFO state is realized.
Now, we examine electron configurations in the AFO state.
Around the lower boundary $U'/D\sim 2$,
the empty and double occupied states alternatively appear for both lattice and orbital,
as shown in Figs.~\ref{AFO}(a) and~\ref{AFO}(b).
Therefore, in this AFO state, the total orbital moment $\Delta n$ is almost zero, which is in stark contrast to the smaller $U'$ case, where
metallic or excitonic state is realized with finite $\Delta n$.
When increasing $U'$, the electron occupancy slightly changes in the sublattice 1,
while drastic change is observed in the other.
In the sublattice 2, the electron number in the orbital $a$ ($b$) smoothly
decreases (increases) beyond $U'/D\sim 2.5$.
This means that the total electron number in orbital $b$ increases continuously
in the AFO phase, which is accompanied by the increase of $\Delta n$
with increasing $U'$.
At last, the AFO phase changes to the band insulating phase, where the orbital $b$ is fully occupied in both sublattices around $U'/D=4.0$, via the continuous phase transition.
In fact, the AFO order parameter monotonically decreases and vanishes at $U'/D=4.0$, as shown in Fig.~\ref{AFO}(c).

\begin{figure}[tb]
  \includegraphics[width=8cm]{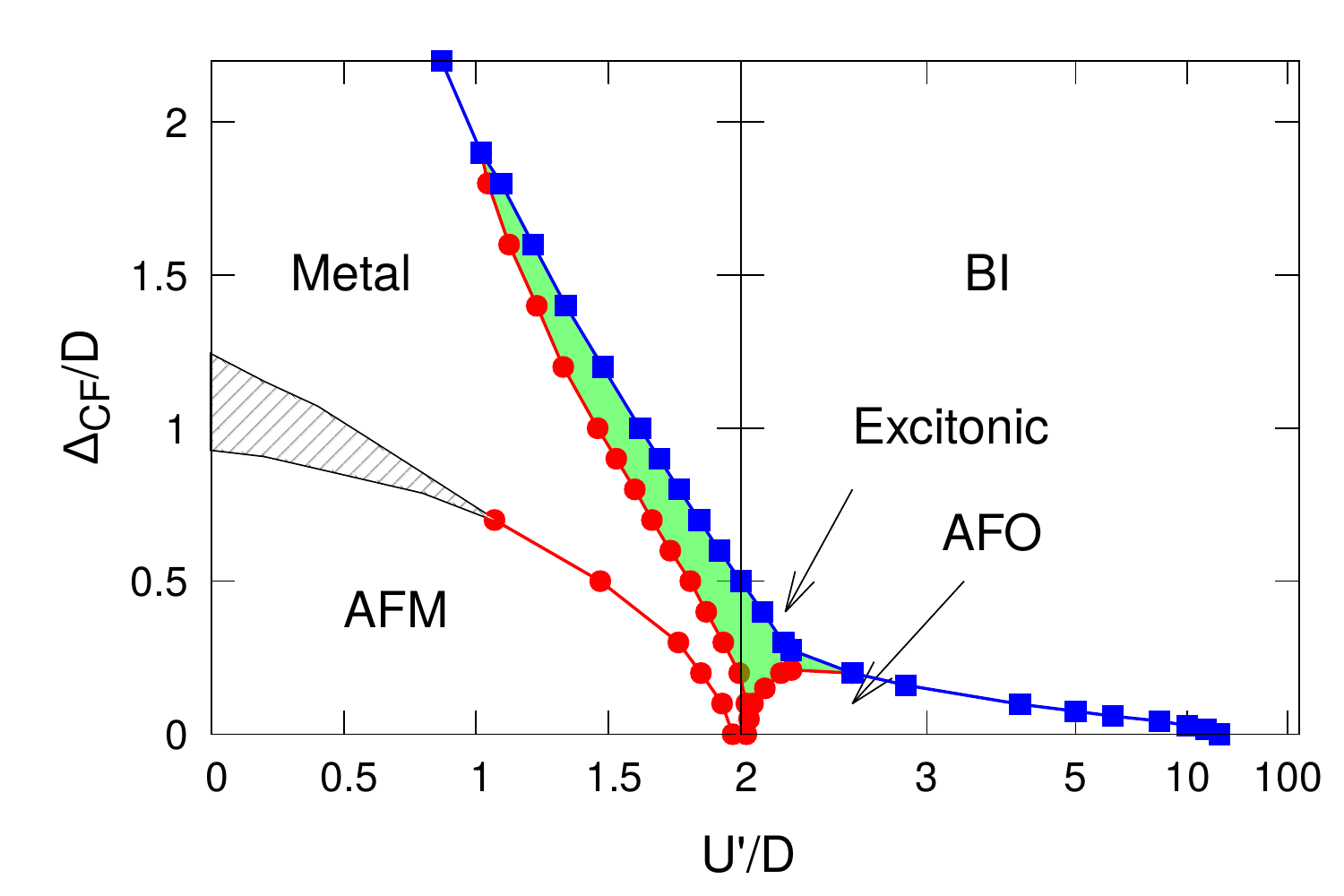}
  \caption{
    Phase diagram for the two-orbital Hubbard model with
    $U/D=2.0$ at the temperature $T/D=0.02$.
    In the shaded area, we could not obtain the converged solutions, but
    the incommensurate state should be realized.
  }
  \label{DV}
\end{figure}
Performing similar calculations,
we obtain the phase diagram at the temperature $T/D=0.02$,
as shown in Fig.~\ref{DV}.
When $U'=0$, the system is reduced to two independent single-band Hubbard models
with the different chemical potentials $\mu_s=\pm \Delta_{\rm CF}/2$.
It is known that the AFM state is realized
around particle-hole symmetric condition~\cite{1367-2630-11-8-083022}.
In fact, we find that
the AFM state is realized in the case with small $\Delta_{\mathrm{CF}}$ and $U'<U$,
as shown in Fig.~\ref{DV}.
The introduction of $\Delta_{\rm CF}$ is regarded as the effect of the self-doping between the two orbitals.
Therefore, $\Delta_{\rm CF}$ is expected to destabilize the AFM in the present system.
In the intermediate $\Delta_{\rm CF}$ region shown as the shaded area,
the staggered magnetization and particle number slowly change in the DMFT iterations
and we could not obtain the converged solutions.
This may imply the existence of the incommensurate magnetic ordered state,
which is consistent with the results for
the doped single-band Hubbard model~\cite{1367-2630-11-8-083022}.
The large $\Delta_{\mathrm{CF}}$ stabilizes the metallic state
until the band insulator is realized.
It is also found that the excitonic phase is realized
in the vicinity of the band insulating phase,
as shown in Fig.~\ref{DV}.
This is consistent with the fact that
the excitons are formed when electrons are almost occupied in the valence band
and low electron density appear in the conduction band.
From this consideration, it is expected that the excitonic state is stable against
other perturbations as far as the system is close to the band insulator.
These results are essentially the same as those
in the previous works~\cite{PhysRevB.85.165135}.

On the other hand, different behavior appears in the $U'>U$ case.
We find that the AFO state is stable rather than the excitonic state,
as shown in Fig.~\ref{DV}.
This should be explained by considering the strong coupling limit $U'\rightarrow \infty$.
The large interorbital interaction realizes
the empty and doubly occupied states in two orbitals,
which leads to doubly degenerate states in each site.
This degeneracy is lifted by the hopping and crystalline electric field
to realize the AFO and BI states, respectively.
Thus, we expect that it is hard to realize the excitonic state in the limit.
Namely, the AFO ordered state should be stabilized in the small $\Delta_{\rm CF}$ region
since the energy gain is given as $t^2/U'$.
When its energy gain smears due to thermal fluctuations,
the metallic (Mott insulating) state is realized
at a small $\Delta_{\rm CF}\neq 0$ ($\Delta_{\rm CF}=0$) region.
What is the most important is that the AFO state is realized
even when the electron density is fractional, as discussed above.
This is in contrast to the AFM state stabilized only around half filling.
Therefore, in the large $U'$ case,
the excitonic state is less stable than the AFO state.

\begin{figure}[tb]
  \begin{center}
    \includegraphics[width=8cm]{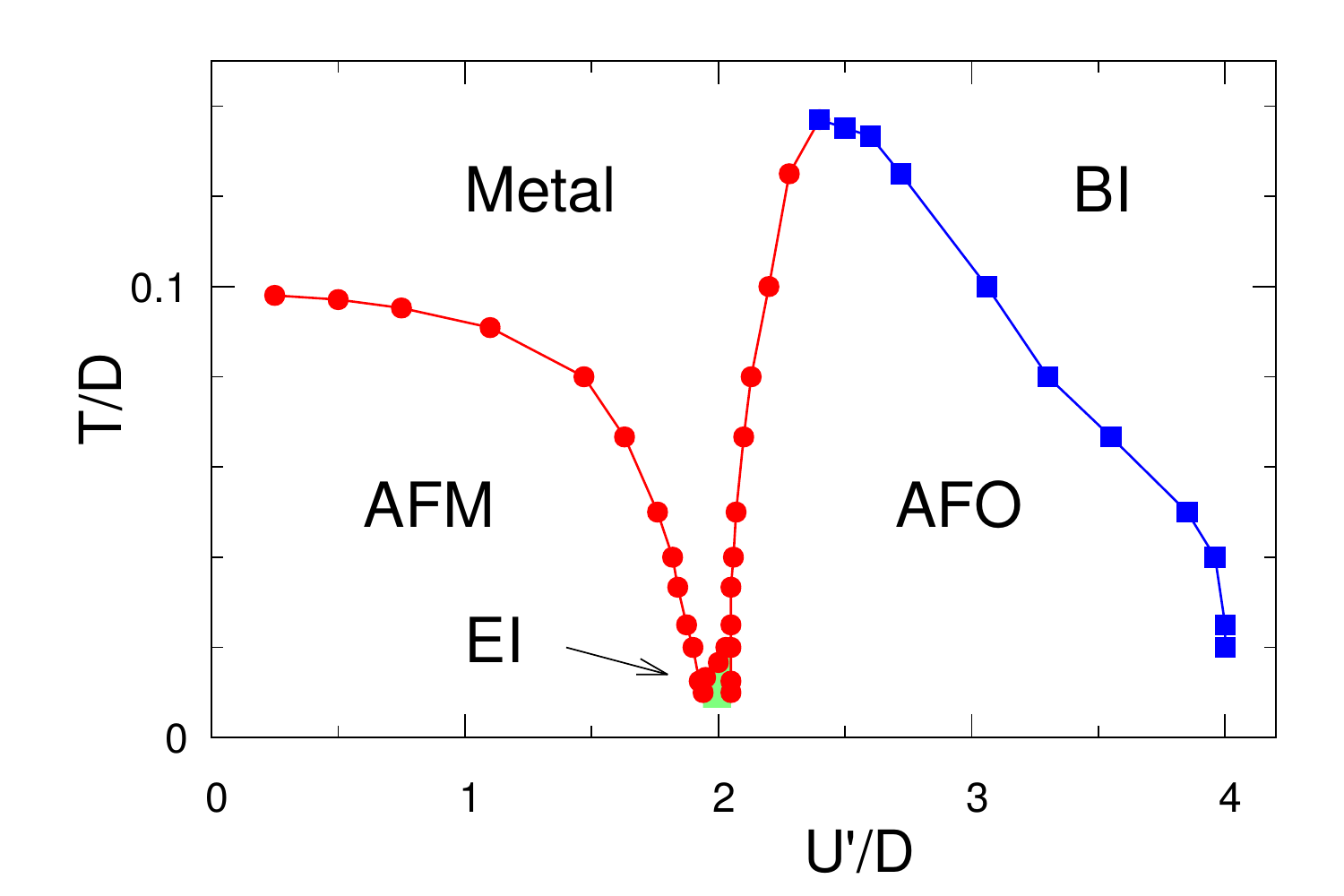}
    \caption{The phase diagram in the two-orbital Hubbard model
      with $\Delta_{\mathrm{CF}}/D=0.1$ and $U/D=2.0$.
    }
    \label{TU}
  \end{center}
\end{figure}
Figure~\ref{TU} shows the finite-temperature phase diagram
with $\Delta_{\rm CF}/D=0.1$.
The AFM and AFO states are widely realized in the case with $U>U'$ and $U<U'$,
respectively.
We also find that these phase boundaries rapidly approach zero around $U=U'$,
which should originate from the fact that
the spin and orbital fluctuations are equally enhanced
and correlated metallic state is realized down to the temperature
$T/D\sim 0.02$~\cite{Koga2band1,Koga2band2}.
We find the excitonic phase around $U=U'$
at lower temperatures.
The above results suggest that the excitonic state is realized due to the enhancement of both spin and orbital fluctuations.

Here, we briefly comment on the competition against
the $s$-wave superconducting (SC) state.
It is known that this SC state is stabilized in the geometrical frustrated
two-orbital model~\cite{capone2002strongly,PhysRevB.91.085108,PhysRevLett.118.177002,Hoshino_Werner_2016,Ishigaki1} and
 the fcc fullerene based compounds are possible candidates.
However, the SC state is stabilized only by orbital fluctuations, leading to
relatively low critical temperatures.
Therefore, we could not find the $s$-wave superconducting phase
in the phase diagram.

\section{Effect of the Hund coupling}\label{sec:Hund}
In this section, we study the stability of the ordered states
against the Hund coupling $J$ under the condition $U=U'+2J$.
In the case with finite $J$,
the excitonic state with parallel spin is distinct from
one with antiparallel spin.
Namely, in the positive (negative) Hund coupling case,
the excitonic state with antiparallel (parallel) spin is stabilized
due to the Ising Hund coupling as mentioned before.
Therefore, calculating the corresponding order parameter in the system,
we discuss the stability of the excitonic state.

\begin{figure}[tb]
  \begin{center}
    \includegraphics[width=8cm]{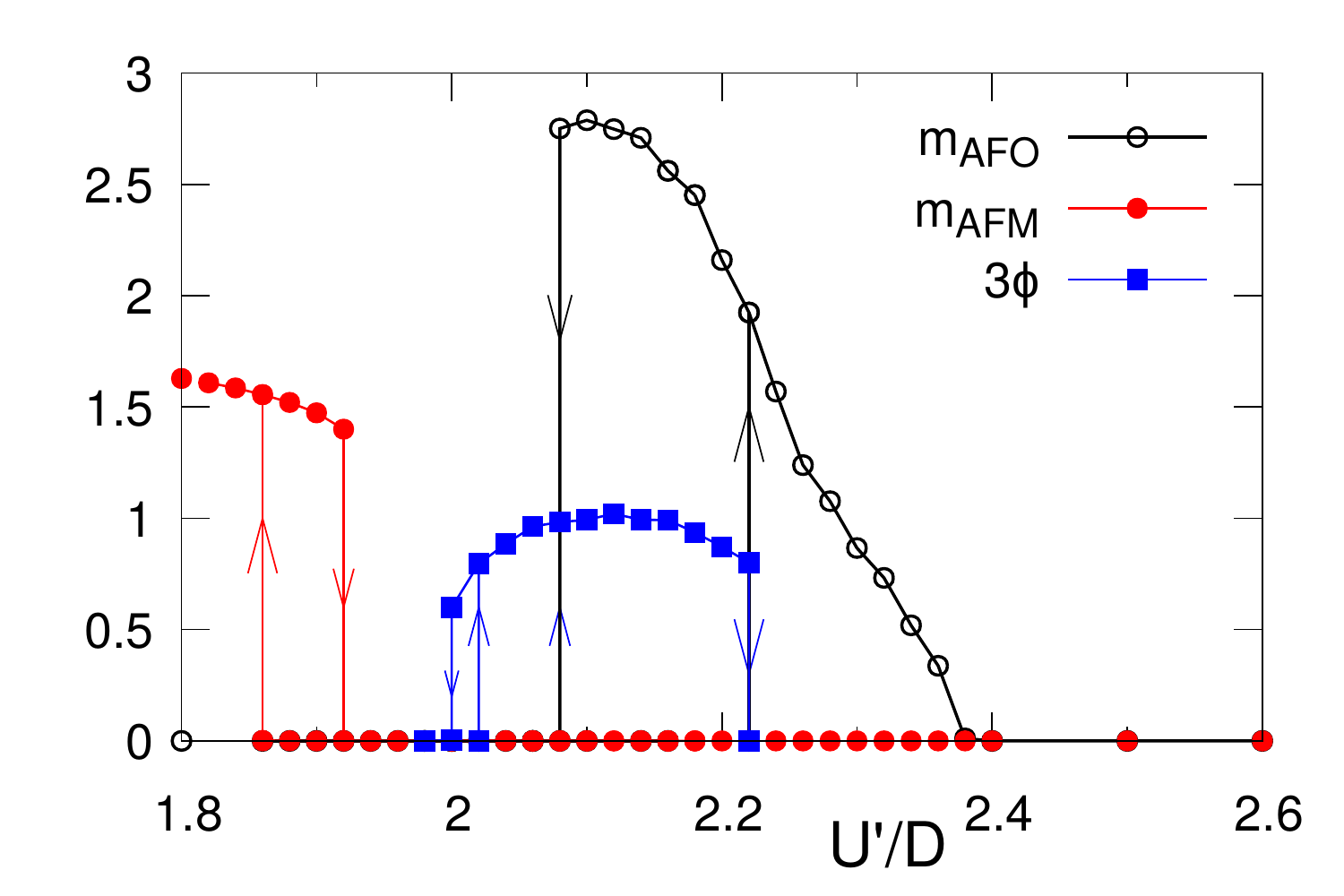}
    \caption{
      Order parameters as a function of the interorbital interaction
      in the two-orbital system with $U/D=2.0$ and $\Delta_{\mathrm{CF}}/D=0.2$
      under the condition $U=U'+2J$ at the temperature $T/D=0.03$.
      }
    \label{Delta0.2}
  \end{center}
\end{figure}
Figure~\ref{Delta0.2} shows the interorbital interaction dependence of the order parameters
in the system with small $\Delta_{\rm CF}$ at the temperature $T/D=0.02$.
When $U'/D\lesssim 1.9$, the interorbital interaction is smaller than the intraorbital interaction, and
the AFM state with the staggered magnetization $m_{AFM}$ is realized.
On the other hand, in the case $U'/D\sim 2.3$, the large interorbital interaction stabilizes
the AFO state.
Around the symmetric condition $U/D=U'/D=2$,
the magnitude of the Hund coupling is relatively small.
Then, the correlated metallic and excitonic states appear
due to large fluctuations in spin and orbital sectors.
Therefore, low energy physics is essentially the same as the system without the Hund coupling,
discussed in the previous section.

\begin{figure}[tb]
 \begin{center}
 \includegraphics[width=8cm]{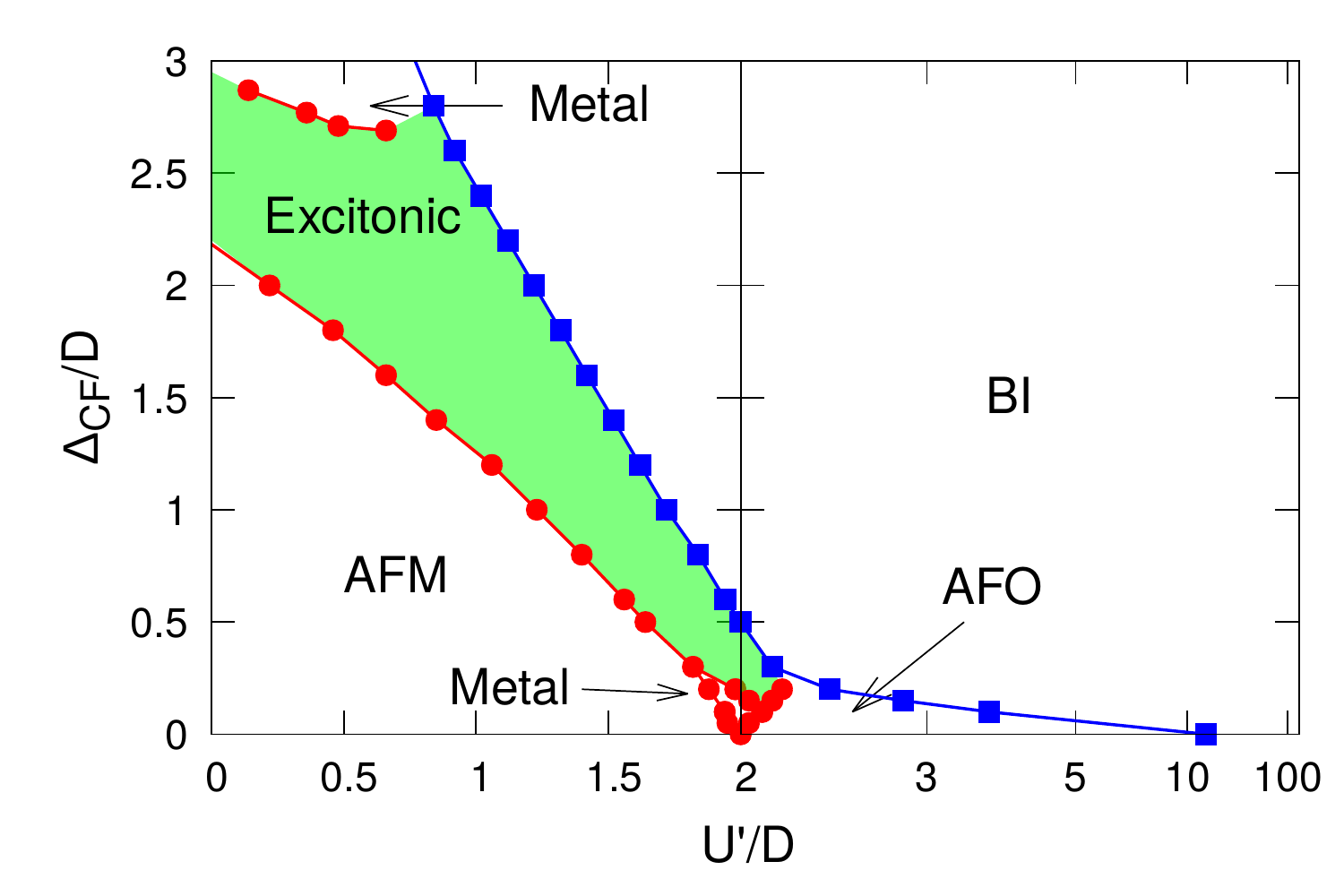}
 \caption{
   Finite temperature phase diagram in the system with $U/D=2.0$ and $U=U'+2J$.
   Solid circles and squares represent the first-order and second-order transition points.
   In the shaded area with $U'>2D$ ($U<2D$),
   the antiparallel-spin (parallel-spin) excitonic state is realized.
 }
 \label{UVtri}
 \end{center}
 \end{figure}

On the other hand, in the large $\Delta_{\rm CF}$ region,
the excitonic state appears away from the particle-hole symmetric condition $U=U'$,
where the effect of the Hund coupling becomes relevant.
Figure~\ref{UVtri} shows the phase diagram in the two-orbital system
at the temperature $T/D=0.02$.
In the case $U'<U$, the excitonic state with the antiparallel spins is more stable
than the metallic state.
In fact, the excitonic state is widely stabilized in the finite temperature phase diagram
in comparison with the phase diagram shown in Fig.~\ref{DV}.
Then metallic states are separated by the excitonic state to smaller and larger $\Delta_{\rm CF}$ regions.
The AFM state is widely realized in the phase diagram since
the Hund coupling between two orbitals enhances magnetic moments.
On the other hand, the AFO state is hardly affected by the Hund coupling.
Therefore, the region of the AFO state is almost unchanged.

\section{Summary}
We have studied the half-filled two-orbital Hubbard model with the crystalline electric field in the wide parameter region including negative Hund coupling case.
This model is analyzed using dynamical mean-field theory with the continuous-time quantum Monte Carlo simulations.
We have clarified that an excitonic phase appears due to the competition between the spin and orbital fluctuations in the case where the intra-orbital interaction is close to the inter-orbital one.
We have also found that, by introducing the Hund coupling, the excitonic state is realized in the wider region in the phase diagram whereas the antiferroorbital phase boundary is almost unchanged.

\section*{Acknowledgments}
Parts of the numerical calculations were performed
in the supercomputing systems in ISSP, the University of Tokyo.
This work was supported by Grant-in-Aid for Scientific Research from
JSPS, KAKENHI Grant Nos. JP18K04678, JP17K05536 (A.K.),
JP16K17747, JP16H02206, JP18H04223 (J.N.).
The simulations have been performed using some of
the ALPS libraries~\cite{alps2}.

\bibliographystyle{jpsj}
\bibliography{./refs}

\end{document}